\documentclass[twoside,twocolumn,english,preprintnumbers,showpacs,showkeys,secnumarabic,amssymb,amsmath,amsfonts,pre]{revtex4-1}
\usepackage[T1]{fontenc}
\usepackage[latin9]{inputenc}
\usepackage{amsmath}
\usepackage{graphicx}
\usepackage{amssymb}
\usepackage{esint}

\makeatletter
\@ifundefined{textcolor}{}
{%
 \definecolor{BLACK}{gray}{0}
 \definecolor{WHITE}{gray}{1}
 \definecolor{RED}{rgb}{1,0,0}
 \definecolor{GREEN}{rgb}{0,1,0}
 \definecolor{BLUE}{rgb}{0,0,1}
 \definecolor{CYAN}{cmyk}{1,0,0,0}
 \definecolor{MAGENTA}{cmyk}{0,1,0,0}
 \definecolor{YELLOW}{cmyk}{0,0,1,0}
 }

\@ifundefined{definecolor}
 {\@ifundefined{definecolor}
 {\usepackage{color}}{}
}{}
\usepackage{dcolumn}

\makeatother

\usepackage{babel}

\makeatother

\usepackage{babel}

\makeatother

\usepackage{babel}

\begin{document}

\title{Self-organization, Pattern Formation, Cavity Solitons and Rogue Waves  \\
in Singly Resonant Optical Parametric Oscillators}

\author{Gian-Luca Oppo}
\email{g.l.oppo@strath.ac.uk}

\author{Alison M. Yao}
\author{Domenico Cuozzo}

\affiliation{ICS, SUPA and Department of Physics, University of Strathclyde, Glasgow
G4 0NG, Scotland, U.K.}

\begin{abstract}
Spatio-temporal dynamics of singly resonant optical parametric oscillators with 
external seeding displays hexagonal, roll and honeycomb patterns, optical turbulence, 
rogue waves and cavity solitons. We derive appropriate mean-field equations with a sinc$^2$ 
nonlinearity and demonstrate that off-resonance seeding is necessary and responsible 
for the formation of complex spatial structures via self-organization. We compare this 
model with those derived close to the threshold of signal generation and find that 
back-conversion of signal and idler photons is responsible for multiple regions 
of spatio-temporal self-organization when increasing the power of the pump field.

\end{abstract}



\maketitle

\section{\label{Intro} Introduction.}
Transverse pattern formation, autosolitons 
and cavity solitons have been the subject of intense research 
in nonlinear optics in the last two decades since their original predictions 
\cite{lugiato87,dalessandro91,rosanov88,tlidi94,firth96}. 
Unlike in other fields of science, transverse patterns and dissipative solitons find 
useful applications in photonics such as optical memories, delay lines and optical registers 
\cite{ackemann09}. Cavity solitons counterparts in the propagation direction have also been 
shown to generate passive mode-locking in fiber lasers \cite{grelu12}.

Formation of transverse spatial structures in quadratic nonlinear cavities was predicted 
first in optical parametric oscillators (OPOs) \cite{oppo94a,devalcarcel96} and later extended 
to second harmonic generation \cite{etrich97,lodahl99}. Early predictions in OPOs were confined 
to the degenerate case where signal and idler fields have the same frequency. Experimental 
evidence of pattern formation was indeed found in triply resonant degenerate OPOs close to 
the confocal cavity configuration \cite{ducci01} and via conical emissions \cite{shelton01,peckus05}. 
Confirmation of the predictions of \cite{oppo94a} was provided in a broad-aperture degenerate 
OPOs in a plane-mirror mini-cavity \cite{peckus07}. Degenerate OPOs also display phase domain 
dynamics and dark-ring cavity solitons \cite{oppo01}. Finally, OPO models for non-degenerate 
Type-II cases in doubly or triply resonant cavity configurations have also been shown to display 
self-organization and pattern formation \cite{oppo94b,staliunas95,longhi96,sanchez97,santagiustina02}.

Transverse instabilities in the case of non-degenerate, singly resonant OPOs (SROPOs), where the 
signal field is the only resonated field in an optical cavity, have been less discussed in the 
literature. On the theoretical side pattern formation in SROPOs is expected to replicate results 
of the complex Ginzburg-Landau laser case \cite{staliunas95}. On the experimental side cw SROPO 
configurations are notoriously difficult to operate because of high oscillation thresholds 
(typically several watts) in common birefringent crystals \cite{yang93}. Quasi-phase matching 
in periodically poled materials has, however, considerably reduced operation thresholds of cw SROPOs 
\cite{bosenberg96} allowing for diode \cite{klein00} and fiber \cite{henderson06} laser pumping 
for spectroscopy applications. A major advantage of cw SROPOs is that their wide tunability is monotonic 
and not affected by mode jumps typical of doubly or triply resonant configurations.

In this paper we investigate the formation and dynamics of transverse structures in SROPOs. We first 
derive a mean-field model in section \ref{MFL} where the nonlinearity is of sinc$^2$ form in agreement 
with early studies of SROPO steady states emissions \cite{kreuzer69,brunner73,brunner77}. The analysis 
builds on approaches that describe and integrate the propagation equations inside the OPO crystal 
\cite{rosencher02,phillips10} by considering transverse effects and by carefully separating the 
mean-field and close-to-threshold approximations. The final model equations are capable of describing 
transverse pattern formation in the presence of pump depletion, signal-idler recombination and 
external seeding close to the signal frequency. External seeding proves to be of fundamental 
importance for transverse structures in SROPOs since, in its absence, changes of the cavity length 
are compensated by changes in the signal (and idler) frequency thus nullifying the common mechanism 
of Turing pattern formation in off-resonant optical systems \cite{lugiato87,oppo09}. 

In section \ref{PWSS} plane-wave steady states and their stability are analyzed in the SROPO models 
with external seeding, close to and far from threshold. These studies confirm that no pattern formation 
should be expected without a detuned external seed. Analytical expressions for the location in the 
parameter space of the loss of stability of homogeneous solutions to spatially modulated structures 
are then provided in section \ref{TUPF}. The thresholds for pattern formation when changing the seeding 
intensity are then compared with those obtained from numerical integration of the SROPO dynamical 
equations with excellent agreement. Section \ref{OTCS} investigates when spatially periodic spatial 
structures break down to either optical turbulence for small seeding intensities or to cavity solitons 
for large pump and seeding intensities. Optical turbulence is demonstrated to be the mechanism which
generates rogue waves in the spatio-temporal evolution of the output fields. Finally, bright and dark 
cavity solitons are found in multistable configurations with localized hexagonal and honeycomb patterns.

\section{\label{MFL} Mean-field models.}

We consider parametric down conversion in a $\chi^{(2)}$ crystal of length $L$ 
at perfect phase matching, a condition that can also describe the average effect of 
quasi-phase matching in periodically poled crystals. In this case the propagation of 
the pump, signal and idler fields in the crystal along the $z$ direction are described 
by \cite{BoydBook}:
\begin{eqnarray}
\partial_z E_0 + \frac {n_0}{c} \,\,\, \partial_t E_0 &=& 
\frac{i}{2k_0} \nabla^2 E_0 - \alpha  E_1 E_2 \nonumber \\
\partial_z E_1 + \frac {n_1}{c} \,\,\, \partial_t E_1 &=& 
\frac{i}{2k_1} \nabla^2 E_1 + \mu \alpha E_0 E_2^*  \label{PropEqu} \\
\partial_z E_2 + \frac {n_2}{c} \,\,\, \partial_t E_2 &=&
\frac{i}{2k_2} \nabla^2 E_2 + \nu \alpha E_0 E_1^* . \nonumber
\end{eqnarray}
where $E_j$ with $j=0,1,2$ are the slowly varying amplitudes of pump, signal and idler fields,
respectively, with wave-numbers $k_j=n_j \Omega_j/c$ and $\nabla^2$ is the transverse Laplacian 
operator along the $x$ and $y$ directions perpendicular to the propagation axis $z$. The 
frequency constraint $\Omega_0 = \Omega_1 + \Omega_2$ is rewritten as $\mu + \nu = 1$ where 
$\Omega_1 = \mu \Omega_0$, $\Omega_2 = \nu \Omega_0$ and the effective
coupling parameter $\alpha$ is given by
\begin{eqnarray}
\label{alpha}
\alpha = \frac {4 \pi \Omega_0 \chi^{(2)}} {n c} 
\end{eqnarray}
where $\chi^{(2)}$ is the second order susceptibility of the crystal, $n=n_0=n_1=n_2$ is the 
common refractive index of the three waves that guarantees phase matching and $c$ is the speed 
of light in vacuum. 

\begin{figure}[tbp]
\begin{center}
\includegraphics[scale=.75]{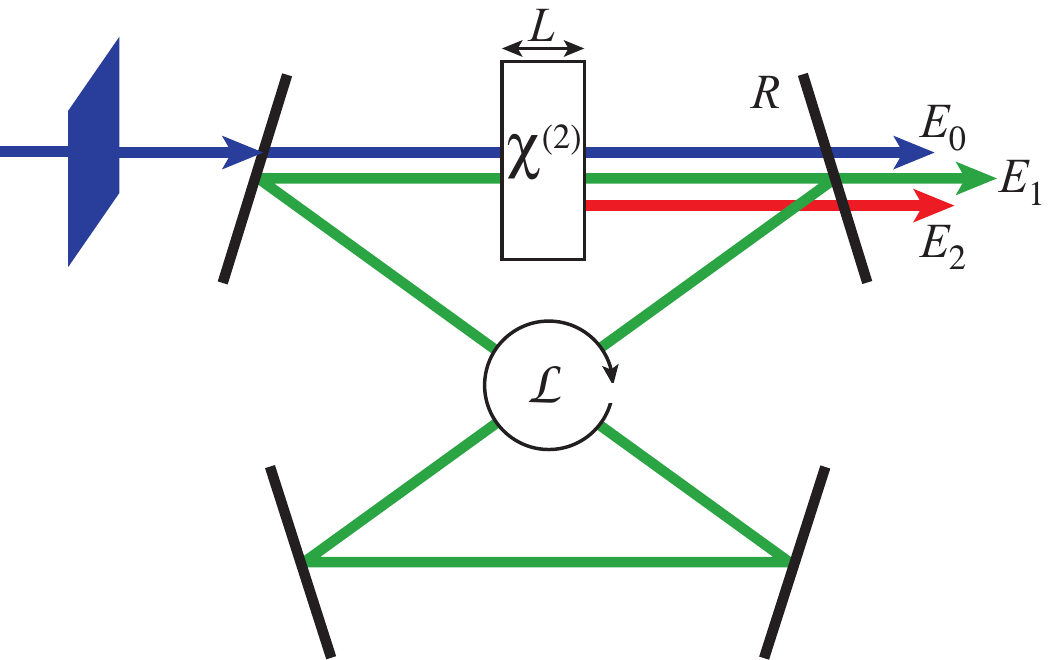}
\end{center}
\caption{(Color online) Schematic diagram of a SROPO cavity of length $\cal{L}$ with a single 
partially reflecting mirror $R$ and containing a parametric down-conversion crystal of length $L$.}
\label{fig:SROPO}
\end{figure}

We assume that the parametric down conversion crystal is contained in an optical cavity of length 
$\cal{L}$ where the signal field is the only one to be resonated (see Figure \ref{fig:SROPO}). The 
steps involved in taking the mean-field approximation are the same as those reported in \cite{santagiustina02} 
although in the SROPO case there is only one resonated field. The final equation for the normalized 
signal field reads as:
\begin{eqnarray}
\label{MFL2}
\tau \,\, \partial_{t'} E_1 &+& L \,\, \partial_{z} E_1 = 
- \gamma E_1 - i \delta E_1 + i a \nabla^2 E_1 \nonumber \\
&+& \mu \alpha L E_0 E_2^* + \sqrt{2 \gamma} E_{IN} \, . 
\end{eqnarray}
where we have introduced 
\begin{equation}
t' = t - \left( \frac{{\cal L} - L}{c} \right) \frac {z}{L}
\end{equation}
and the parameters
\begin{eqnarray}
\label{NPAR}
\tau &=& \frac {{\cal L} + (n - 1) L}{c} ; \;\;\;\;\;\;\;\;\;\;\;
\gamma = \frac{1-R} {2} ;  \\
\delta &=& \frac {\omega_c - \Omega_1} {c} {\cal L}; \;\;\;\;\;\;\;\;\;\;\;\;\;\;\;\;\;
a = \frac {{\cal L}} {2 k_1} \, .
\end{eqnarray}
Here, $R$ is the output mirror reflectivity, $\omega_c$ is the frequency of the longitudinal cavity 
mode closest to the signal frequency $\Omega_1$ and $E_{IN}$ is a complex input field of frequency 
$\omega_{IN}$ close to $\Omega_1$, normally known as the seeding.

The usual mean-field limit procedure requires high reflectivity $R$ and involves an expansion in 
longitudinal Fourier modes and the requirement that all terms, including the nonlinear one, are 
independent of the longitudinal variable $z$. The $z-$variation per pass of the resonated signal 
field, $E_1$, can be neglected when it is affected by the average of the propagation of the pump 
and idler waves along the crystal \cite{phillips10}, i.e.
\begin{equation}
\label{average}
E_1 = \frac{1}{L} \int_0^L E_0(z) \, E_2^*(z) dz \, .
\end{equation} 
To obtain an explicit dependence of pump and idler fields along the direction of propagation 
we consider the first and third equations of the system (\ref{PropEqu}) and neglect diffraction 
in the crystal:
\begin{eqnarray}
d_{z} E_0(z) &=& - \alpha  E_1 E_2(z) \label{eqn:s1} \\
d_{z} E_2(z) &=&  \nu \alpha E_0(z) E_1^*  \label{eqn:s2}
\end{eqnarray}
where the signal amplitude $E_1$ is now independent of $z$. By taking the second derivative 
of (\ref{eqn:s1}) and using (\ref{eqn:s2}), one obtains
\begin{equation}
d_{z}^2 E_0(z) = - (\nu \alpha^2 I_1) E_0(z)
\end{equation}
which shows that the pump field oscillates along the propagation direction with a frequency 
that depends on the signal intensity $I_1$. Integrating this equation we find
\begin{equation}
\label{e0}
E_0(z) = A_0 \cos \left( \alpha \sqrt{\nu I_1} z \right) \, ,
\end{equation}
where $A_0$ is the amplitude of the pump field at the entrance of the crystal \cite{phillips10}.
From (\ref{eqn:s1}),
\begin{equation}
\label{e2}
E_2(z) = - \frac{1}{\alpha E_1} d_{z} E_0 = A_0 E_1^* \sqrt{\frac{\nu}{I_1}} 
\sin \left( \alpha \sqrt{\nu I_1} z \right)
\end{equation}
in agreement again with \cite{phillips10}. 

We can now calculate the spatial average (\ref{average}):
\begin{equation}
\label{av2}
\frac{1}{L} \int_0^L E_0(z) \, E_2^*(z) dz = |A_0|^2 \frac{E_1}{2 \alpha L I_1} 
\sin^2 \left( \alpha L \sqrt{\nu I_1} \right) \, 
\end{equation}
and insert it into (\ref{MFL2}):
\begin{eqnarray}
\label{MFL3}
\tau \,\, \partial_{t'} E_1 &+& L \,\, \partial_{z} E_1 = 
- \gamma E_1 - i \delta E_1 + i a \nabla^2 E_1 \\
&+& \mu |A_0|^2 \frac{E_1}{2 I_1} 
\sin^2 \left( \alpha L \sqrt{\nu I_1} \right) + \sqrt{2 \gamma} E_{IN} \, . \nonumber
\end{eqnarray}
By expanding in longitudinal Fourier modes and retaining only the longitudinal mode 
closest to $\Omega_1$, corresponding to $\partial_{z'} E_1 = 0$, we finally obtain:
\begin{eqnarray}
\label{MFL4}
\partial_{t'} E_1 &=& \kappa [ -(1+i\theta) E_1 
+ i {\hat a} \nabla^2 E_1 \nonumber \\
&+& \mu |A_0|^2 \frac{E_1}{2 \gamma I_1} 
\sin^2 \left( \alpha L \sqrt{\nu I_1} \right) + {\hat E}_{IN} ] 
\end{eqnarray}
where
\begin{eqnarray}
\kappa &=& \frac{\gamma}{\tau} = \frac{\gamma c}{{\cal L}+(n-1)L} \, ; 
\;\;\;\;\;\;\;\;\;
{\hat a} = \frac{a}{\gamma} =  \frac{{\cal L}}{2k_1\gamma} ; \nonumber \\
\theta &=& \frac{\delta}{\gamma}= \frac{(\omega_c-\Omega_1){\cal L}}{c \gamma} \, ; 
\;\;\;\;\;\;\;
{\hat E}_{IN} = \sqrt{\frac{2}{\gamma}} \, E_{IN}\, .
\end{eqnarray}
Finally, we renormalize the transverse space variables $x$ and $y$ by dividing them by 
$\sqrt{{\hat a}}$, the time variable by multiplying it by $\kappa$ and the field 
amplitudes according to
\begin{eqnarray}
E &=& \alpha L \sqrt{\nu} \,\, E_1 \, ; 
\;\;\;\;\;\;\;\;\;\;\;
|E_0|^2 = |A_0|^2 \frac{\mu \nu \alpha^2 L^2 }{2 \gamma} \nonumber \\
E_{IN} &=& \alpha L \sqrt{\nu} \,\, {\hat E}_{IN}
\end{eqnarray}
to obtain
\begin{eqnarray}
\label{SROPO}
\partial_{\tau} E = \partial_{\kappa t'} E &=& E_{IN} -(1+i\theta)E \\ 
&+& |E_0|^2 \frac{E}{I} 
\sin^2 \left( \sqrt{I} \right) + i \nabla^2 E \, . \nonumber
\end{eqnarray}
The analysis of Eq. (\ref{SROPO}) is the main focus of the research presented here. It will be 
referred to as the \textit{sinc$^2$ model} since $\sin^2 ( \sqrt{I} ) / I = \mathrm{sinc}^2 ( \sqrt{I} )$. 

We note that in SROPO configurations the frequency of the signal field, $\Omega_1$, is tuneable 
by corresponding changes of the idler frequency, $\Omega_2$, while maintaining the energy 
conservation condition $\Omega_0=\Omega_1+\Omega_2$. This means that with no external 
seeding ($E_{IN}=0$) the detuning $\theta$ is also zero since the SROPO tunes its signal 
frequency to the closest longitudinal cavity mode $\omega_c$. With an external seeding different 
from zero and detuned with respect to the cavity, it is advantageous to consider the 
external frequency $\omega_{IN}$ as reference and introduce
\begin{equation}
\theta = \frac{(\omega_{IN}-\Omega_1){\cal L}}{c \gamma} \, .
\end{equation}
Under these conditions $E_{IN}$ should be considered to be real and equation (\ref{SROPO}) 
remains unchanged.

\begin{figure}[!th]
\begin{center}
\includegraphics[width=0.52 \textwidth]{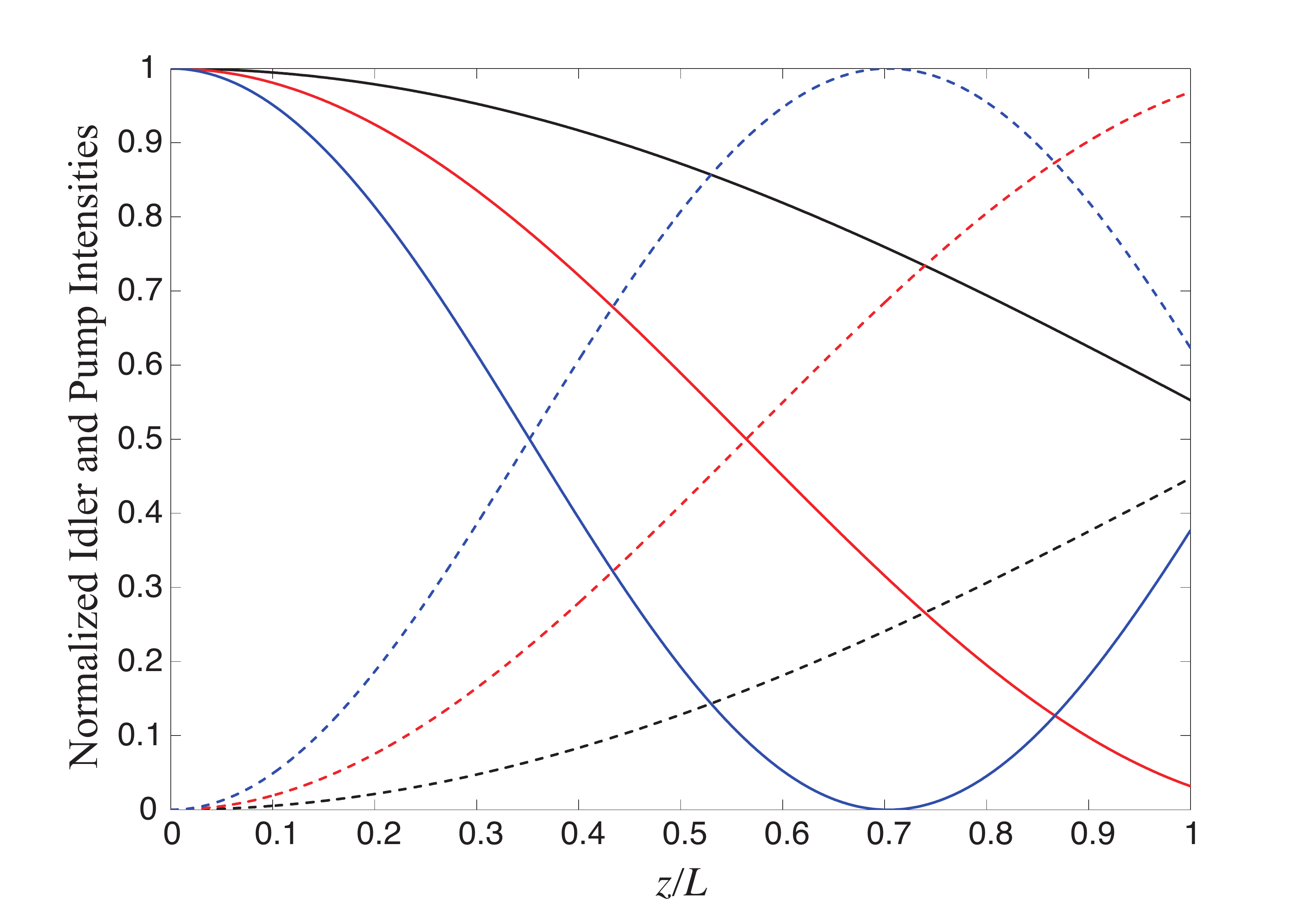}
\end{center}
\caption{\label{fig:CP} (Color online) Pump intensity (solid lines) and idler 
intensity (dashed lines) in the SROPO crystal for $|E_0|^2=1.2$ (black lines), 
$|E_0|^2=2.0$ (red lines) and $|E_0|^2=8.0$ (blue lines). The intensities are 
normalized to the input pump values $|E_0|^2$. The propagation distance is normalized 
to the crystal length.}
\end{figure}

It is interesting to investigate the behavior of the pump and idler fields inside the OPO
crystal as provided by Eqs. (\ref{e0}) and (\ref{e2}). Figure \ref{fig:CP} shows the pump
and idler intensities during propagation for three sample values of $|E_0|^2$, namely 1.2,
2.0 and 8.0. While at $|E_0|^2=1.2$ (black lines) the changes of pump and idler per 
pass are limited, for $|E_0|^2=2.0$ (red lines) and $|E_0|^2=8.0$ (blue lines) they 
are substantial. In particular, full pump depletion and substantial 
back-conversion of signal and idler fields into the pump are clearly visible in Figure 
\ref{fig:CP} for $|E_0|^2=8.0$. In the SROPO case these phenomena are not incompatible 
with the mean field approximation and are at the base of the sinc$^2$ nonlinearity of model 
(\ref{SROPO}). The mean field approximation implies that the signal intensity remains 
almost constant with respect to its input value during propagation in the $\chi^{(2)}$ medium 
with large changes taking place over several cavity round-trips. No such constrains apply to 
pump and idler fields as shown in Fig. \ref{fig:CP}.

Note that the cos$^2$ and sin$^2$ nature of the pump and idler intensities, respectively, guarantees 
conservation of the energy density in every point along the SROPO crystal. Energy conservation 
in turn guarantees the validity of the Manley-Rowe relations about the variations of the
energy densities $N_i$ per field along the crystal:
\begin{equation}
\label{MR}
\frac{dN_0}{dz} = - \frac{dN_1}{dz} - \frac{dN_2}{dz} 
\end{equation}
since $dN_1/dz=0$. 
These facts are a-posteriori confirmations that the physical processes described in Eqs. (\ref{PropEqu}) 
are compatible with the application of the mean-field limit to the signal field even for large 
values of the pump and seeding intensities.

\subsection{\label{CTA} The close-to-threshold approximation.}
Close to the signal generation threshold it is possible to obtain partial differential 
equations in the mean-field limit where the nonlinear terms are in a polynomial form 
and thus easier to analyze. The scaling of the mean field limit requires that the 
nonlinear coefficient 
per pass, $\alpha L$, has to be of the order of the mirror transmittivity, $1-R$. This 
implies that the argument of the $\sin^2$ term in equation (\ref{av2}) may become large for 
large signal intensities without breaking the mean-field conditions. Close to threshold, 
however, the signal intensity satisfies $I_1<1$ and the $\sin^2$ term can be approximated 
by a power expansion. In this case pump and idler display small changes per pass across 
the crystal meaning that pump depletion and back-conversion do not take place in a single 
pass. Equations (\ref{e0}) and (\ref{e2}), however, tell us that while the pump can be 
approximated to first order to a constant value $A_0$, the idler has to grow along $z$ 
from its initial value. This is in agreement with previous analysis below threshold where 
the important noise term is associated with the idler fluctuations at the entrance of the 
crystal \cite{cuozzo11}. In the case of SROPOs close to threshold, we can approximate 
$E_0$ and $E_2$ in (\ref{e0}) and (\ref{e2}) with 
\begin{eqnarray}
\label{e0sc}
E_0(z) &\approx& A_0 \left( 1 - \frac{\nu I_1 \alpha^2 z^2}{2} \right) \\
E_2(z) &\approx& A_0 E_1^* \left( \nu \alpha z - \frac{\nu^2 I_1 \alpha^3 z^3}{6} \right).
\end{eqnarray}
By using these expressions to evaluate the average (\ref{average}) one obtains:
\begin{eqnarray}
\label{av4}
& &\frac{1}{L} \int_0^L E_0(z') \, E_2^*(z') dz' \nonumber \\
& &\approx  \frac{\nu \alpha L |A_0|^2 E_1}{2} \left( 1 - 
\frac{\nu \alpha^2 L^2 I_1}{3} \right) \, .
\end{eqnarray}
By repeating the same steps of the mean-field limit as described in the previous subsection
we obtain:
\begin{equation}
\label{SROPOb}
\partial_{\tau} E = E_{IN} - (1+i\theta)E + |E_0|^2 \left( E-\frac{E I}{3} \right) 
 + i \nabla^2 E 
\end{equation}
which describes the spatio-temporal behaviour of the SROPO close to threshold in the presence 
of an external seeding $E_{IN}$ and will be referred to as the \textit{cubic model}.

\section{\label{PWSS} Plane wave steady-states}
\begin{figure}[tbp]
\begin{center}
\includegraphics[width=0.5 \textwidth]{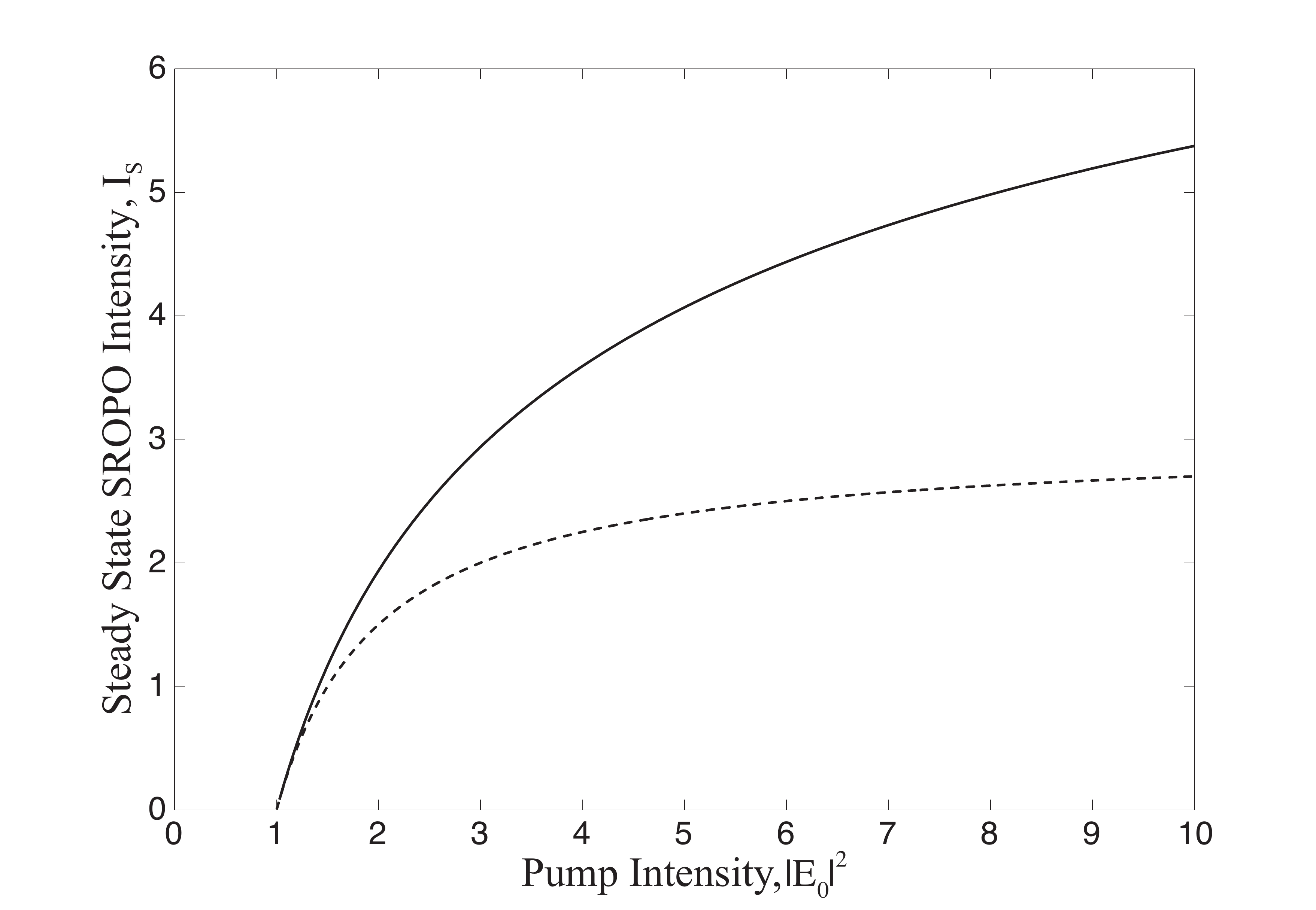}
\end{center}
\caption{\label{fig:SS} Intensity of the SROPO steady-state for the sinc$^2$ model (solid line) 
and the cubic approximation (dashed line) with increasing pump intensities for $E_{IN}=0$ 
and $\theta=0$. All variables are dimensionless.}
\label{ssSROPO}
\end{figure}

As mentioned in section \ref{MFL}, when there is no external seeding, $E_{IN}=0$, the 
detuning is zero since the SROPO automatically adjusts its frequency to the closest cavity 
resonance. The plane wave steady-state intensities, $I_s$, are implicit for the sinc$^2$ 
model (see \cite{kreuzer69,brunner73,brunner77}) and explicit for the cubic model:
\begin{eqnarray}
\label{ssnoin}
I_s &=& |E_0|^2 \sin^2 \left( \sqrt{I_s} \right) \\
I_s &=& 3(|E_0|^2-1)/|E_0|^2 \nonumber
\end{eqnarray}
The steady-state signal intensity of the SROPO as a function of the pump intensity, $|E_0|^2$, 
is shown in Figure \ref{fig:SS} for the sinc$^2$ model (solid line) and the cubic 
approximation (dashed line). These are trivially complemented by the zero-intensity state 
that is stable below threshold, $|E_0|^2<1$, and unstable above. In the cubic case the 
stationary intensity above threshold asymptotes to the value $3$ for large pump intensities 
and is always stable. The steady-state curve for the sinc$^2$ model, on the other hand, becomes 
multivalued at large values of the input pump intensity ($|E_0|^2 > 20$, not shown here) 
\cite{brunner73,brunner77}. 
Here, however, we are interested in values of the pump intensity below $10$, as these are more 
realistic with respect to present state-of-the-art of broad area SROPO realisations. In this regime 
it is possible to prove that, above threshold, the non-zero steady-state intensities in the sinc$^2$ 
model are also stable \cite{brunner77}. Note that when comparing the sinc$^2$ and the cubic models, 
there is a substantial difference between their steady-state intensities even below $|E_0|^2 = 2$. 
At twice above threshold this difference becomes considerable and the close to threshold (cubic) 
model has to be discarded.

Analogously to lasers, the field phase is decoupled from the steady-state equations and 
is affected by fluctuations and drift processes. When there is external seeding, $E_{IN} > 0$, 
the phase of the SROPO locks to that of the external beam, depending on 
the magnitude of the detuning $\theta$ and the input intensity. Such behavior strongly differs 
from that of $E_{IN}=0$. In the case of $E_{IN} \neq 0$ the steady-state intensities are given 
by 
\begin{equation}
\label{SSwSeed}
E_{IN}^2 = I_s \left[ \left ( 1 - |E_0|^2 f(I_s) \right )^2 + \theta^2 \right] 
\end{equation}
where
\begin{eqnarray}
f(I_s) &=&  \mathrm{sinc}^2(\sqrt{I_s}) \\ 
\nonumber \\
f(I_s) &=&  1 - I_s/3 
\end{eqnarray}
for the sinc$^2$ and cubic models, respectively. The steady-state curves of the SROPO intensity 
versus the input intensity become S-shaped, a behavior typical of injected optical systems, as 
shown in section \ref{TUPF}.

For the cubic model without diffraction it is possible to obtain analytical results.
For example, for $|\theta|<(|E_0|^2 - 1 )/\sqrt{3}$ the plane-wave steady-state curves are S-shaped,  
and the positions of the turning points $[(E_{IN}^2)^{-},I_s^{-}]$ and $[(E_{IN}^2)^{+},I_s^{+}]$ 
can be determined by finding the maxima and minima of (\ref{SSwSeed}):
\begin{equation}
I_s^{\pm} = \frac{2 (|E_0|^2 - 1) \pm \left( (|E_0|^2 - 1)^2-3\theta^2 \right)^{1/2}}{|E_0|^2} 
\end{equation}
and then using these values in (\ref{SSwSeed}). At resonance, $\theta=0$, the turning points are 
located at:
\begin{eqnarray}
\label{cubeplusminus}
\left[(E_{IN}^2)^+,I_s^+\right] &=& \left[ 0,3 (|E_0|^2 - 1)/|E_0|^2 \right] \\
\left[(E_{IN}^2)^-,I_s^-\right] &=& \left[ 4 (|E_0|^2 - 1)^3 / (9 |E_0|^2), \right . \nonumber \\
& & \left . (|E_0|^2 - 1) / (3 |E_0|^2)\right] \nonumber \, .
\end{eqnarray}
Note that the $+$ turning point at resonance corresponds to the zero seeding case of SROPO 
intensity given by Eq. (\ref{ssnoin}).

\subsection{\label{LSA} Linear stability analysis of the SROPO with seeding.}
\begin{figure}[!tb]
\includegraphics[width=0.52\textwidth]{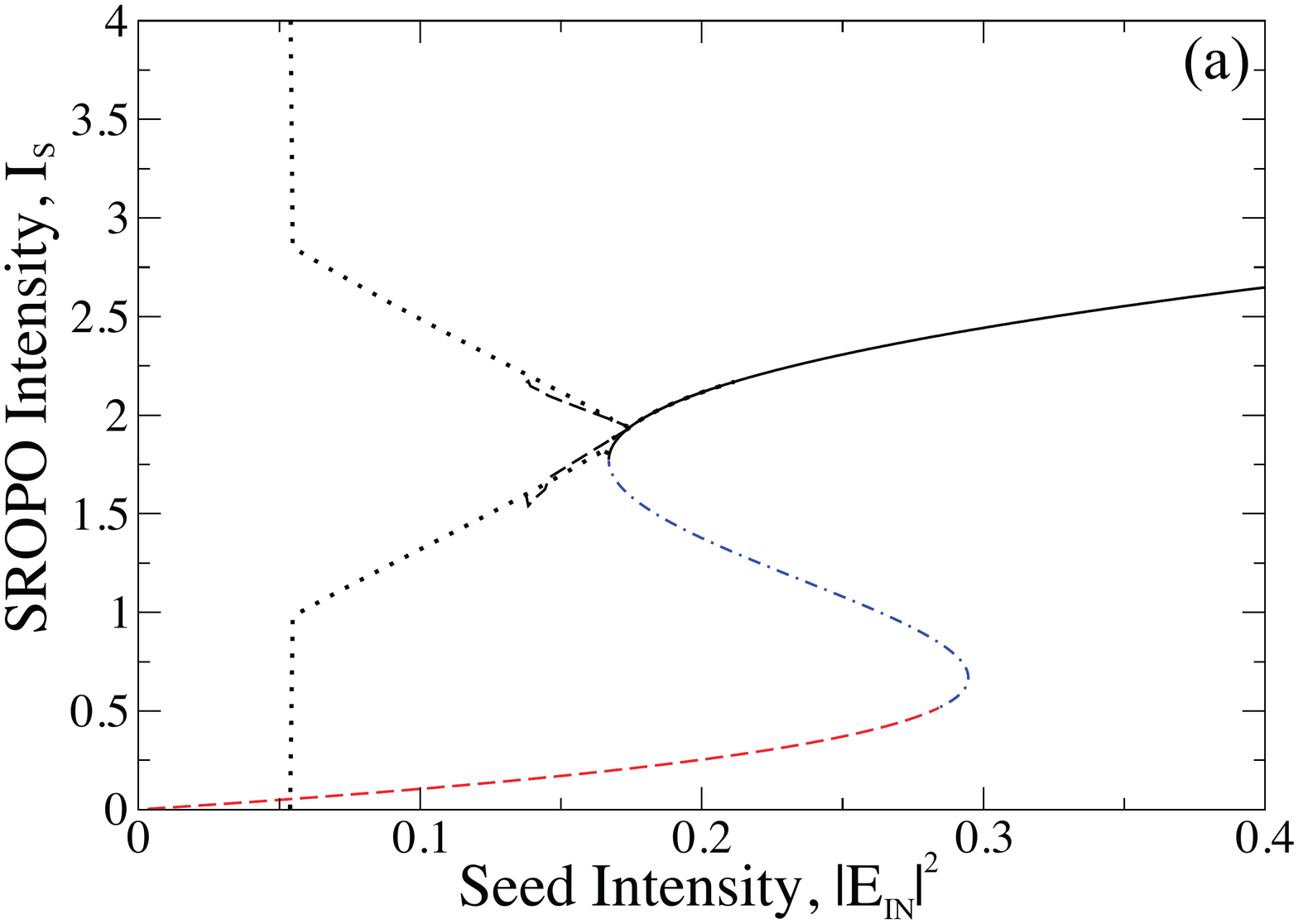} 
\includegraphics[width=0.52\textwidth]{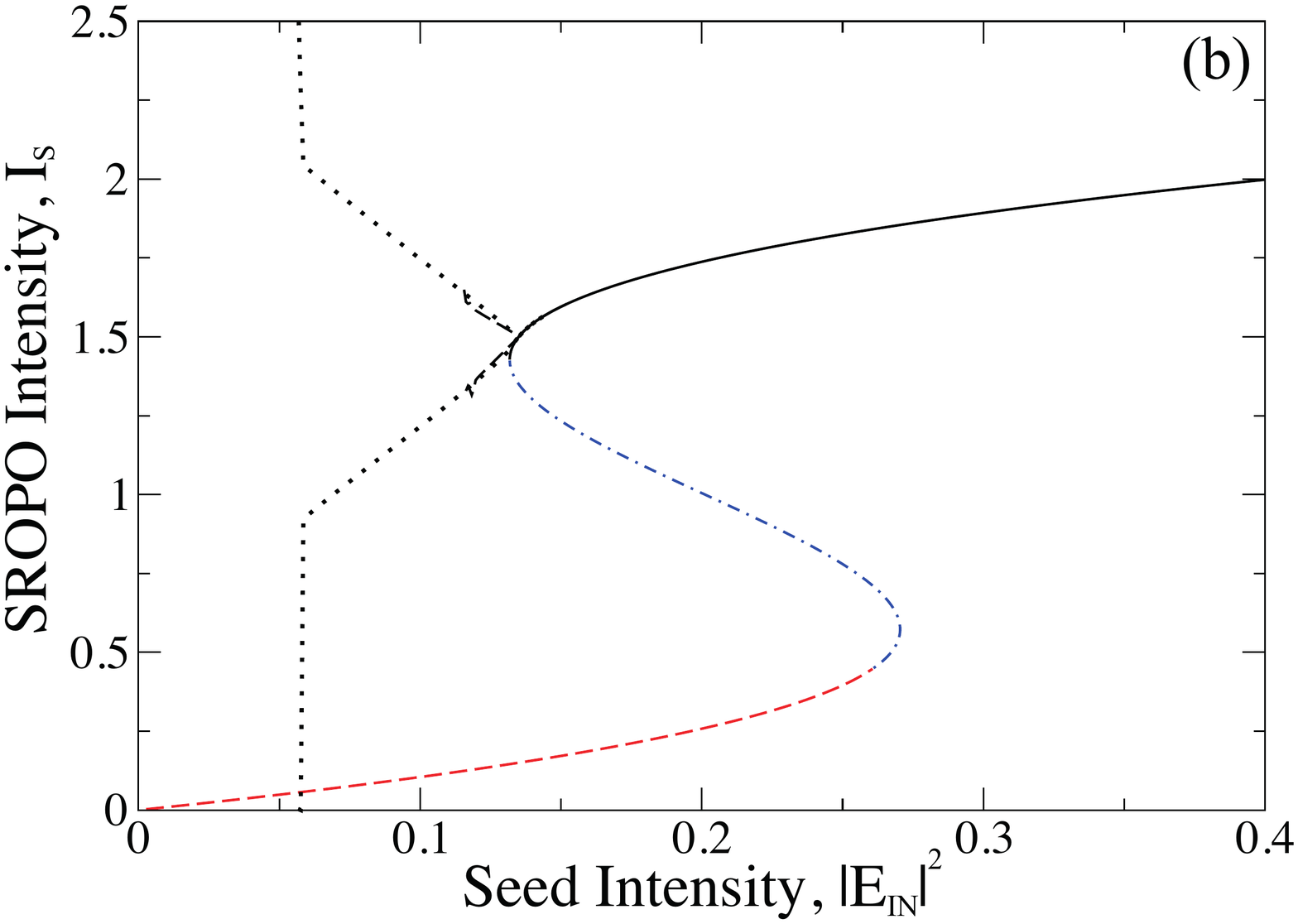} 
\caption{\label{fig:PFe02_2} (Color online) Plane wave steady-state stability and pattern formation 
for (a) the sinc$^2$ model (\ref{SROPO}) of a SROPO and (b) the cubic model (\ref{SROPOb}) of 
a SROPO close to threshold. 
The solid (black), dot-dashed (blue) and dashed (red) 
lines correspond to stable, unstable and Hopf unstable plane wave steady states, respectively. The 
black dotted (black dashed) lines correspond to the minimum and maximum of the intensity of stationary 
hexagonal (roll) patterns. The vertical dotted line corresponds to the instability of hexagons leading 
to optical turbulence. Parameters are $|E_0|^2=2$ and $\theta=-0.3$. All variables are dimensionless.
}
\end{figure}

The linear stability analysis of the steady-states given in the previous section produces two 
stability eigenvalues:
\begin{equation}
\label{evalues}
\lambda_{\pm} = \xi \pm \sqrt{\beta^2 - \theta^2}
\end{equation}
where for the sinc$^2$ model
\begin{eqnarray}
\label{absinc}
\xi &=& |E_0|^2 \mathrm{sinc}(2\sqrt{I_s})-1 \\
\beta &=& |E_0|^2 \frac{\cos(2\sqrt{I_s})+\sqrt{I_s}\sin(2\sqrt{I_s})-1}{2 I_s} \nonumber \\
\end{eqnarray}
and for the cubic model
\begin{eqnarray}
\label{abcube}
\xi &=& |E_0|^2 - 1 - 2 |E_0|^2 I_s / 3\\
\beta &=& - |E_0|^2 I_s / 3 \nonumber \, .
\end{eqnarray}
For the sinc$^2$ model, the stability eigenvalues are implicit functions of the steady-state 
intensity, $I_s$. It is, however, easy to display the stability of the stationary states graphically 
along the S-shaped curves by picking increasing values of $I_s$, evaluating $\lambda_{\pm}$ and 
reporting the stability result on the diagram, as displayed in Figures \ref{fig:PFe02_2} 
and \ref{fig:PFe02_8}. Here black solid lines correspond to two negative real eigenvalues (sinks), 
turquoise solid lines to stable complex eigenvalues (foci), dot-dashed blue lines to at least one 
positive real eigenvalue (saddles or sources) and red dashed lines to complex eigenvalues with 
positive real part (unstable foci). In terms of bifurcations, the intersection of a black solid 
line and a blue dot-dashed line signals a saddle-node bifurcation, while the transition of a 
turquoise solid line into a red dashed line signals a Hopf bifurcation. 

We find that the turning points of the S-shaped curves always correspond to either saddle-node 
(the $[(E_{IN}^2)^+,I_s^+]$ points) or saddle-source (the $[(E_{IN}^2)^-,I_s^-]$  points) 
bifurcations corresponding to a change of sign of one real eigenvalue. For the cubic model this 
fact can be demonstrated analytically. In the lowest branch of the S-curve, the two real eigenvalues 
turn complex (see the red dashed line in Figures \ref{fig:PFe02_2} and \ref{fig:PFe02_8}). 
This means that the lowermost part of the S-curve is Hopf unstable. 
%
%
\section{\label{TUPF} Turing instabilities and pattern formation}

In this section we describe instabilities of the stationary states of the SROPO to transverse 
perturbations due to diffraction with and without external seeding. By moving to the spatial 
Fourier space of the transverse wave-vector $k$ and repeating the linear stability steps of 
the previous section, we obtain the two stability eigenvalues of (\ref{evalues}) but with 
the detuning $\theta$ replaced by 
\begin{equation}
\label{thetak}
\theta_k=\theta + k^2
\end{equation}
which introduces an explicit dependence on the transverse spatial scale. 

We start with the analysis of possible Turing instabilities without external seeding ($E_{IN}=0$) 
and with zero detuning $\theta=0$. In this case, the evaluation of the stability eigenvalues with 
the appropriate factor (\ref{thetak}) is done only at the values of $I_s$ given by (\ref{ssnoin}). 
In the cubic case the eigenvalues reduce to:
\begin{equation}
\lambda_{\pm} = -(|E_0|^2-1) \pm \sqrt{(|E_0|^2-1)^2-k^4} \, .
\end{equation}
The largest eigenvalue has a zero value for the plane-wave case, $k=0$, corresponding to the 
uncoupled phase of the SROPO models without seeding as studied in the previous section. 
For large wave-vectors the eigenvalues can become complex, i.e. one may observe damped oscillations.
However, the presence of diffraction cannot make the real part of the eigenvalues positive which 
means that, for the SROPO alone, there are no spatio-temporal instabilities and hence no pattern 
formation. We obtain the same result for the sinc$^2$ model within the pump intensity ranges studied 
here although the implicit nature of the steady-state (\ref{ssnoin}) requires straightforward 
numerical evaluations of the stability eigenvalues for given wave-vectors $k$. 

We now consider the case of external seeding where the detuning, $\theta$, can be non-zero. 
By using the expressions (\ref{evalues}) with $\theta$ replaced by $\theta_k$ (\ref{thetak}) one 
observes that the transverse wave-vector can destabilise the system only when it counterbalances 
the detuning and that this is most effective when 
\begin{equation}
\label{offresonance}
k^2 = -\theta 
\end{equation}
i.e. the off-resonance mechanism for pattern formation typical of optical systems \cite{lugiato87,oppo94a}. 
We refer to the off-resonance mechanism as Turing pattern formation since it has been demonstrated 
that all the requirements of Turing instabilities are fully satisfied \cite{oppo09}. 

\begin{figure}[!th]
\includegraphics[width=0.58 \textwidth]{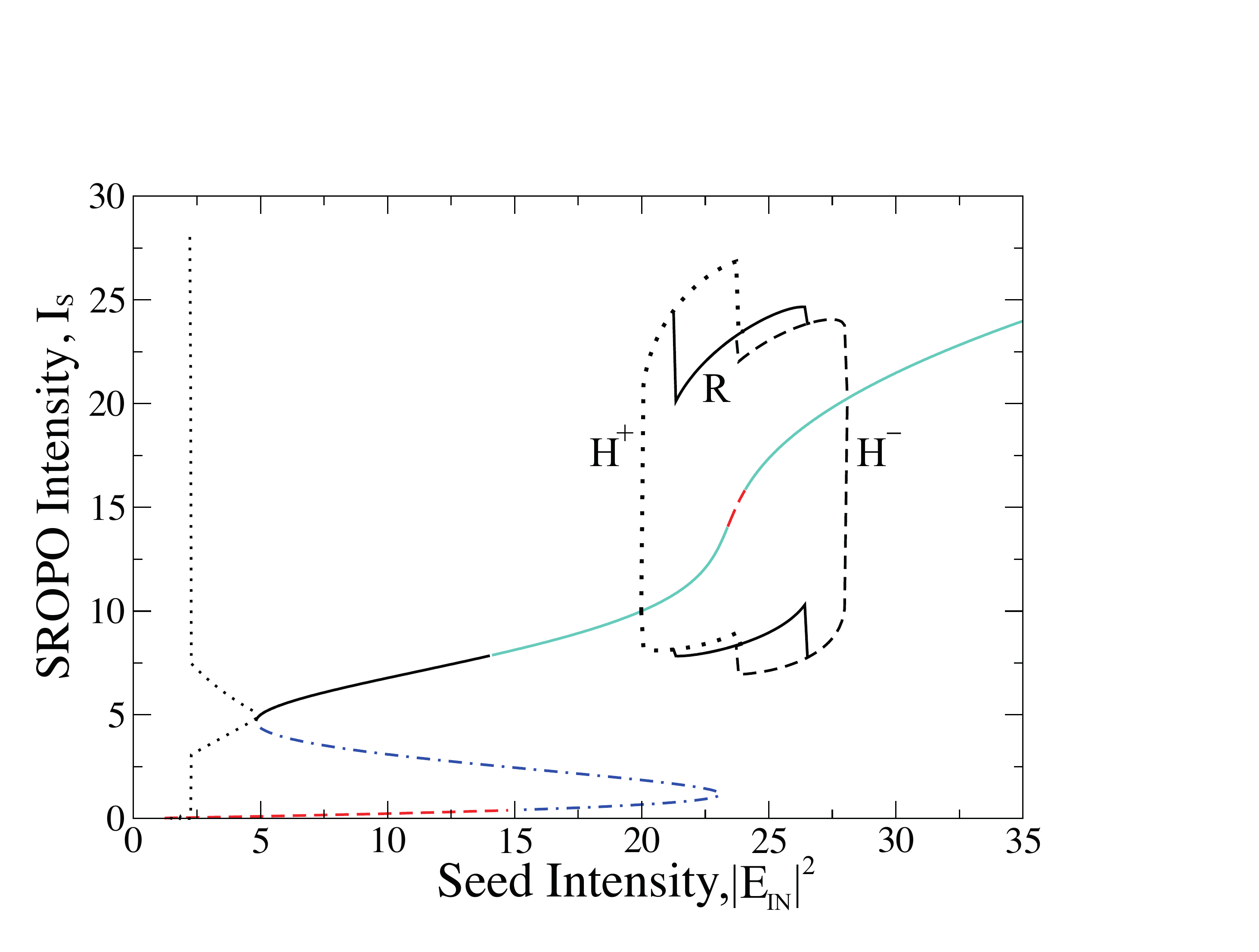} 
\caption{ \label{fig:PFe02_8} (Color online) Same as Fig. \ref{fig:PFe02_2} but for parameter values 
$|E_0|^2=8$ and $\theta=-1$  (all variables are dimensionless). The solid turquoise lines correspond to stable plane wave steady states 
with complex stability eigenvalues. For seed intensities above 20, minima and maxima of the intensity 
of stable hexagonal patterns H$^+$ (dotted lines), of stable roll patterns R (solid lines) and of 
stable honeycomb patterns H$^-$ (dashed lines) are displayed.
}
\end{figure}

The condition (\ref{offresonance}) provides us with the value along the steady-state curves at which we
expect pattern formation to occur, $I_s^c$. This value simply corresponds to the steady-state value of 
the plane wave solution at zero detuning (\ref{ssnoin}) since for $\theta_k=0$ the stability eigenvalues 
(\ref{evalues}) reduce to $\lambda_{\pm} = \xi \pm \beta$, where $\xi$ and $\beta$ are given by 
(\ref{absinc}) for the sinc$^2$ model and (\ref{abcube}) for the cubic model. By tracing a horizontal 
line at the $I_s^c$ value on the diagrams of Figures \ref{fig:PFe02_2} and 
\ref{fig:PFe02_8} one obtains the corresponding value of $|E_{in}^c|^2$ of the seeding intensity where 
the Turing instability takes place.
The bifurcation from the homogeneous states to steady transverse patterns is obtained when decreasing 
the seeded amplitude $E_{IN}$ so that the locked plane wave state progressively approaches the 
upper turning point of the S-shaped steady-state curve (the $[(E_{IN}^2)^+,I_s^+]$ point). 
Before reaching it, the stationary plane wave intensity reaches the value $I_s^c$ and a stationary roll 
pattern is formed supercritically while a hexagonal pattern is formed subcritically in agreement with 
\cite{ciliberto90}. 
This bifurcation scenario is in agreement with early analysis of complex Ginzburg-Landau models in the 
presence of injection \cite{coullet92a,coullet92b}) although our cubic model does not contain diffusion 
or purely imaginary nonlinearities. It is also in remarkable agreement with numerical simulations, as 
demonstrated in section \ref{NUM}.

\begin{figure}[!tb]
\includegraphics[width=0.5\textwidth]{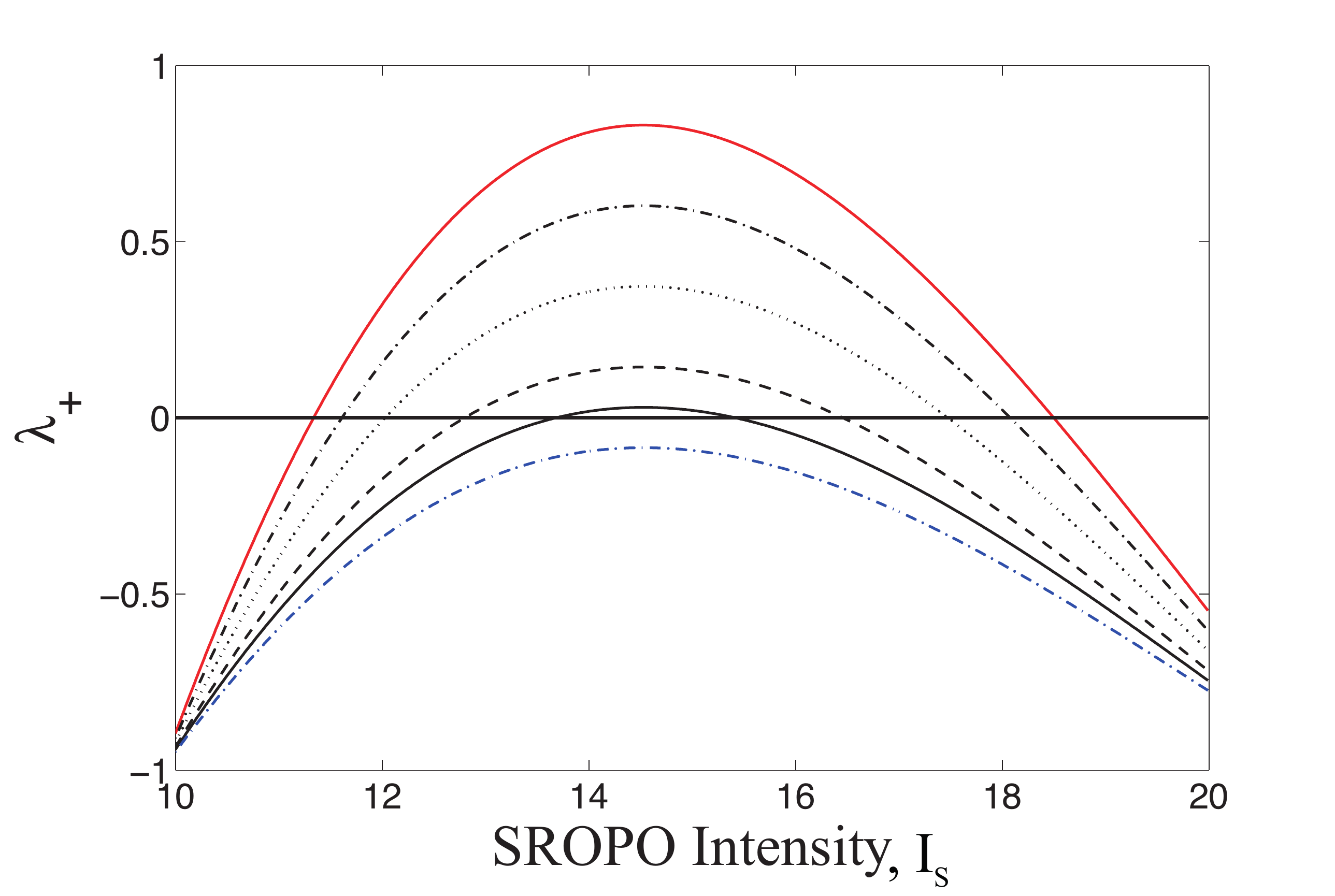} 
\caption{ \label{fig:lambdaplus} (Color online) Stability eigenvalue $\lambda_+$ versus the SROPO 
stationary intensity. Parameters are $\theta=-1$, $|E_0|^2=4$ (lowermost dashed-dotted blue line), 
$|E_0|^2=4.5$ (solid black line), $|E_0|^2=5$ (dashed black line), $|E_0|^2=6$ (dotted black line), 
$|E_0|^2=7$ (uppermost dashed-dotted black line) and $|E_0|^2=8$ (uppermost solid red line). All variables are dimensionless.}
\end{figure}

\begin{figure}[!tb]
\includegraphics[width=0.5\textwidth]{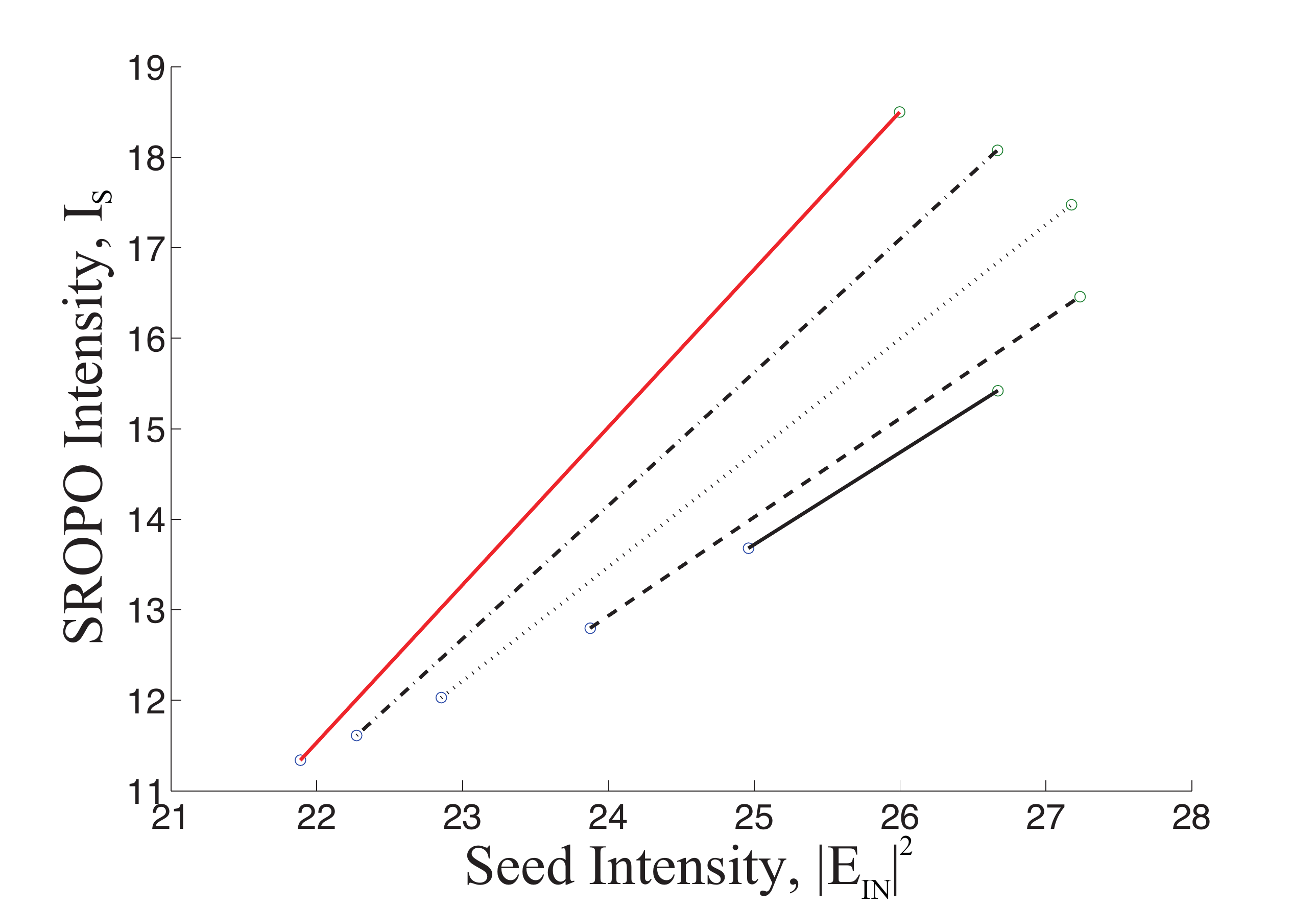} 
\caption{ \label{fig:instabregime} (Color online) Plane wave instability regions to spatial patterns 
in the (seed intensity, SROPO intensity) parameter space. Parameters are $\theta=-1$, 
$|E_0|^2=4.5$ (solid black line), $|E_0|^2=5$ (dashed black line), 
$|E_0|^2=6$ (dotted black line), $|E_0|^2=7$ (uppermost dashed-dotted black line) and $|E_0|^2=8$ 
(uppermost solid red line).  All variables are dimensionless.}
\end{figure}

We have also investigated instabilities of the plane wave to pattern structures for large values of both 
the input pump and the seeding intensity as shown in Fig. \ref{fig:PFe02_8}. These instabilities have no 
counterpart in the close-to-threshold regime and can be estimated analytically by using the stability 
eigenvalues (\ref{evalues}) with (\ref{absinc}) and $\theta_k=0$ for the most unstable wave-vector 
(\ref{offresonance}). Figure \ref{fig:lambdaplus} shows the instability eigenvalue $\lambda_+$ versus 
the stationary SROPO intensity for different values of the input pump $|E_0|^2$. Above a threshold value 
of $|E_0|^2\sim4.37$ (corresponding to a critical value of $I_s=14.5$), there is a range of values of 
the SROPO intensity where the plane wave solution is unstable to spatial patterns. The limit values 
of the SROPO intensity are the zeroes of the $\lambda_+$ curve shown in Figure \ref{fig:lambdaplus} 
with the lower (upper) intersection corresponding to an instability when increasing (decreasing) 
the seeding intensity. 
In Figure \ref{fig:instabregime} we show the plane-wave instability range in the parameter space 
of the SROPO intensity versus the seed intensity for different values of the pump intensity. 
In section \ref{NUM} we show that the bifurcations at the boundaries of the instability ranges are 
subcritical in nature and that there are extended regions of bistability between patterns and stable 
plane waves to support cavity solitons. The ranges displayed in Fig. \ref{fig:instabregime} 
provide a minimum size of the parameter region where pattern formation is expected. For example, the 
plane-wave instability range for $|E_0|^2=8$ evaluated analytically from the stability eigenvalues 
is approximately between $|E_{IN}|^2=22$ and $|E_{IN}|^2=26$ (see Fig. \ref{fig:instabregime}) while 
the numerical simulations find stable patterns between $|E_{IN}|^2=20$ and $|E_{IN}|^2=28$ because 
of subcriticality (see Fig. \ref{fig:PFe02_8}).  
%

\subsection{\label{NUM} Numerical patterns}
\begin{figure}[!tb]
\includegraphics[width=0.45\textwidth]{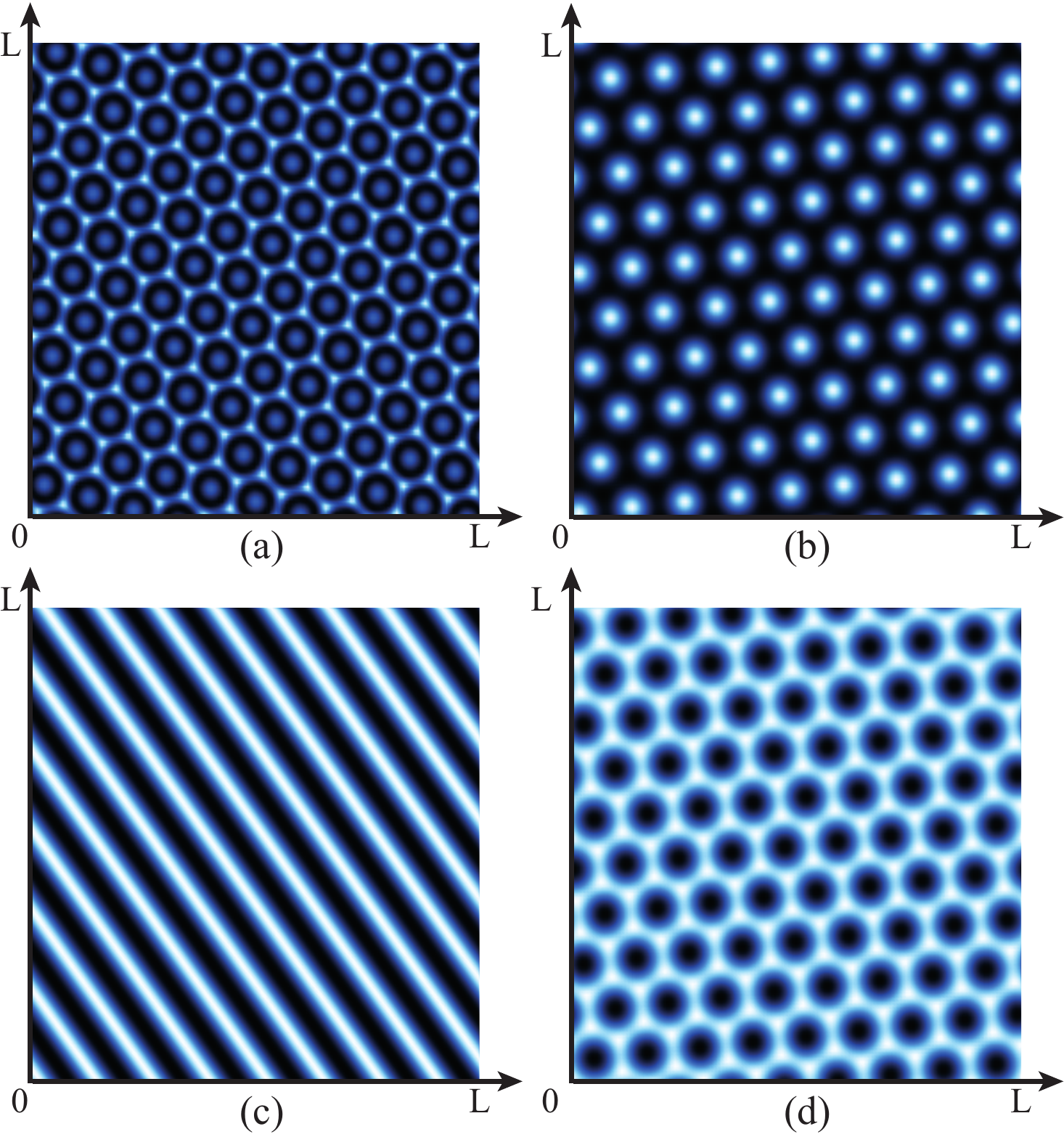} 
\caption{ \label{fig:patterns} (Color online) Intensity of transverse patterns in a SROPO. (a) Hexagons 
for $|E_{IN}|^2=3$. (b) Hexagons for $|E_{IN}|^2=22$. (c) Rolls for $|E_{IN}|^2=24$. (d) Honeycombs 
for $|E_{IN}|^2=27$. Parameters are $|E_0|^2=8$ and $\theta=-1$. All variables are dimensionless.}
\end{figure}

We have first numerically integrated the sinc$^2$ (\ref{SROPO}) and cubic (\ref{SROPOb}) models 
for $|E_0|^2=2$ and $\theta=-0.3$. We have started with relatively large values of the seeding 
amplitude, $E_{IN}=0.45$, where the stable plane-wave solution has been recovered. By progressively 
decreasing $E_{IN}$, a supercritical roll pattern is observed to appear at around 
$E_{IN}=0.424$, $I_s^c = 1.9$ for the sinc$^2$ model and $E_{IN}=0.374$, $I_s^c = 1.5$ for the cubic 
model, in excellent agreement with the theoretical predictions given in section \ref{TUPF}. By further 
decreasing the seeding intensity, the amplitude of the roll pattern increases (see black dashed lines 
in Fig. \ref{fig:PFe02_2} until it merges into a hexagonal structure. Having 
located the hexagonal pattern (see Fig. \ref{fig:patterns} (a) for its transverse intensity structure), 
we have traced it with increasing and decreasing values of the external seeding intensity. 
For small seeding intensities, Figures \ref{fig:PFe02_2} and \ref{fig:PFe02_8} show 
the maximum and minimum intensity of the hexagonal pattern (dotted lines) and show that these change 
linearly with decreasing seeding intensity. The bifurcation back to the steady plane-wave solution is 
subcritical although the regime of sub-critical bistability is very small and difficult to detect on 
the scales of the diagrams. When further decreasing the external seeding, one observes a sudden destabilisation 
of the hexagonal pattern into a region of optical turbulence. The abrupt transition from stable patterns 
to turbulence is clearly displayed in Figures \ref{fig:PFe02_2} and \ref{fig:PFe02_8} 
by the almost vertical line on the right hand side of these diagrams that corresponds to a sudden jump 
in the values of the minima and maxima intensities observed in the transverse section during the turbulent 
evolution.

For larger values of the input pump, $|E_0|^2$, new regions of pattern formation arise in the SROPO with 
seeding in a way similar to what has been described for nascent optical bistability \cite{tlidi94}. These 
new regions can only be observed in the sinc$^2$ model since the cubic model can only 
display stable plane wave solutions for large $|E_0|^2$ and large $E_{IN}^2$. Moreover, the cubic model 
is not accurate away from threshold. For the numerical simulations presented here we have selected the 
value of $|E_0|^2=8$ where the minimum size of the pattern region is more than 15\% of the maximum value 
of the seed intensity in order to guarantee relevance to possible experimental realizations. In Fig. 
\ref{fig:PFe02_8} we present the intensities of the observed patterns together with the steady-state 
plane wave curves for the selected value of $|E_0|^2=8$. 
At low seed intensities the phenomenology is similar to that described for $|E_0|^2=2$ above. However, 
at larger seeding intensities the upper branch of the S-shaped plane-wave steady-state curve suddenly 
increases. The steady-state first develops damped oscillations and then becomes unstable to a Hopf 
bifurcation (see dashed red lines around the seed intensity of 23 in Figure \ref{fig:PFe02_8}). 
Around such bifurcation, a new region of stationary patterns develops. We have identified rolls R 
(solid lines), hexagons H$^+$ (dotted lines) and honeycombs H$^-$ (dashed lines). The intensities of the 
different transverse patterns are displayed in Figure \ref{fig:patterns}. We note that none of the 
patterns observed at large input pumps and seeding intensities are present in the cubic model. Finally, 
pattern bistability is observed between rolls and hexagons and rolls and honeycombs.

\section{\label{OTCS} Optical turbulence, rogue waves and cavity solitons.}
\begin{figure}[!tb]
\includegraphics[width=0.55 \textwidth]{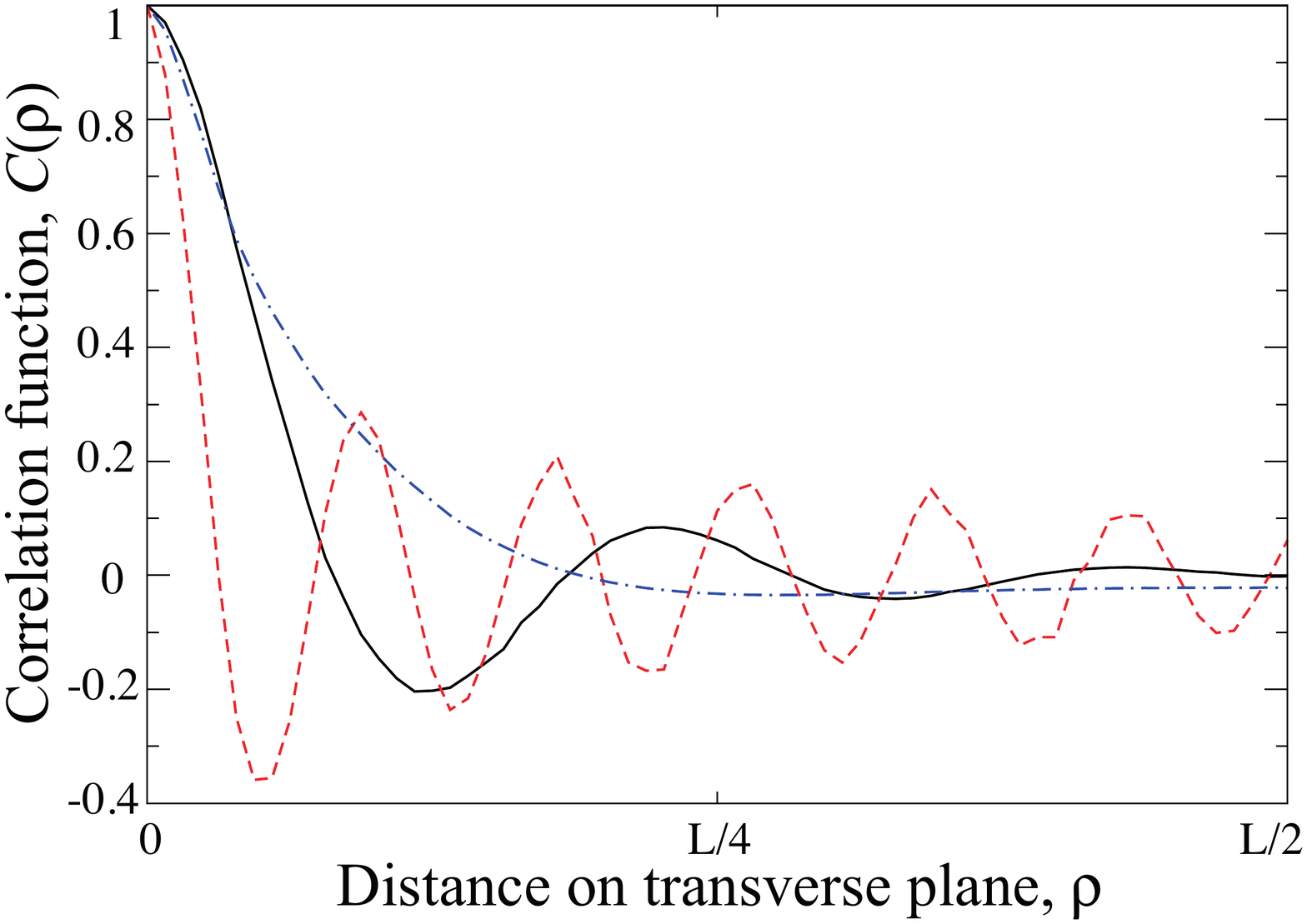}  
\caption{ \label{fig:corr} (Color online) Spatial correlation function $C(\rho)$ as in (\ref{corrfunct}) 
for the hexagonal pattern (red dashed line, $|E_0|^2=2$, $\theta=-0.3$, $|E_{IN}|^2=0.09$), optical turbulence 
close to threshold (black solid line, $|E_0|^2=2$, $\theta=-0.3$, $|E_{IN}|^2=0.04$) and away from threshold 
(blue dot-dashed line, $|E_0|^2=8$, $\theta=-1.0$, $|E_{IN}|^2=2.19$). All variables are dimensionless.}
\end{figure}

When the seeding is small, the input energy is not sufficient to lock the SROPO to the external 
laser. These unlocked regimes are typical of lasers with injected signals \cite{oppo86}. The larger 
the detuning, $\theta$, between the external laser and the SROPO cavity, the larger the seed intensity 
necessary for locking. Since the lower branch of the S-shaped steady-state curves is always Hopf unstable 
for small seeding, one expects to observe dynamical regimes where locking and unlocking alternate in 
space and time. In comparison with purely temporal systems, the presence of transverse degrees of 
freedom elongates the locking region to lower values of the seeding intensity, as displayed in Figures 
\ref{fig:PFe02_2} and \ref{fig:PFe02_8} where stable hexagons are observed well into 
the region where plane wave solutions are unstable. 
As the seeding intensity is decreased, unlocking 
eventually takes place and stable patterns develop defects \cite{ouyang91,coullet92b} that induce 
first phase and then amplitude instabilities. The resulting regime corresponds to optical turbulence since 
one observes a sudden (exponential) decrease of the spatio-temporal correlation function \cite{harkness94}
\begin{eqnarray}
\label{corrfunct}
C(\rho) =  
\frac{{\rm Re}\left[\langle E({\bf r},t) E^*({\bf r}',t) \rangle - 
\langle E({\bf r},t) \rangle  \langle E^*({\bf r},t) \rangle \right]}
{{\rm Re}\left[\langle E({\bf r},t) E^*({\bf r},t) \rangle - 
\langle E({\bf r},t) \rangle  \langle E^*({\bf r},t) \rangle \right]}
\end{eqnarray}
where ${\bf r}$ and ${\bf r}'$ identify separate positions on the transverse plane, $\rho=|{\bf r}-{\bf r}'|$, 
${\rm Re}$ denotes the real part and $\langle \cdot \rangle$ corresponds to temporal averages. 
Such behavior is demonstrated in Figure \ref{fig:corr} 
where the correlation function $C(\rho)$ is calculated for the hexagonal pattern (dashed line), the 
turbulent regimes for $|E_0|^2=2$ (solid line) and $|E_0|^2=8$ (dot-dashed line). Fitting exponentials to the 
correlation functions shows that in the turbulent regimes the correlation length is 
reduced by at least a factor of six. 
In the regime of optical turbulence, large 
variations of the SROPO intensity are observed in both space and time. In Figures \ref{fig:PFe02_2}
and \ref{fig:PFe02_8} we display the range of variation of the SROPO intensity at a 
given time $t$ at the onset of optical turbulence. The wide increase in the maximum SROPO intensity 
when changing the seed strength below the hexagon instability is clearly visible. 

\begin{figure}[!tb]
\includegraphics[width=0.55 \textwidth]{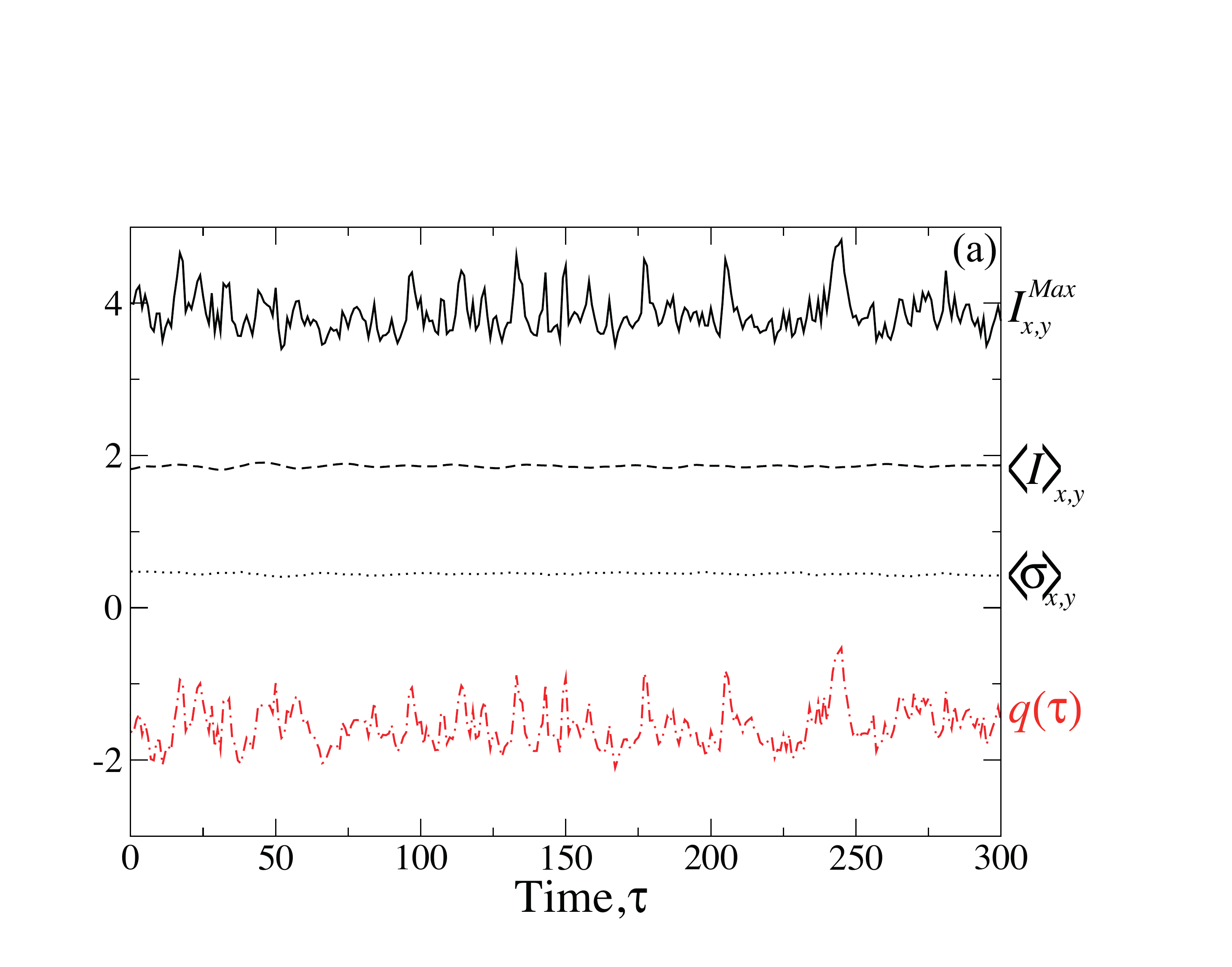} 
\includegraphics[width=0.55 \textwidth]{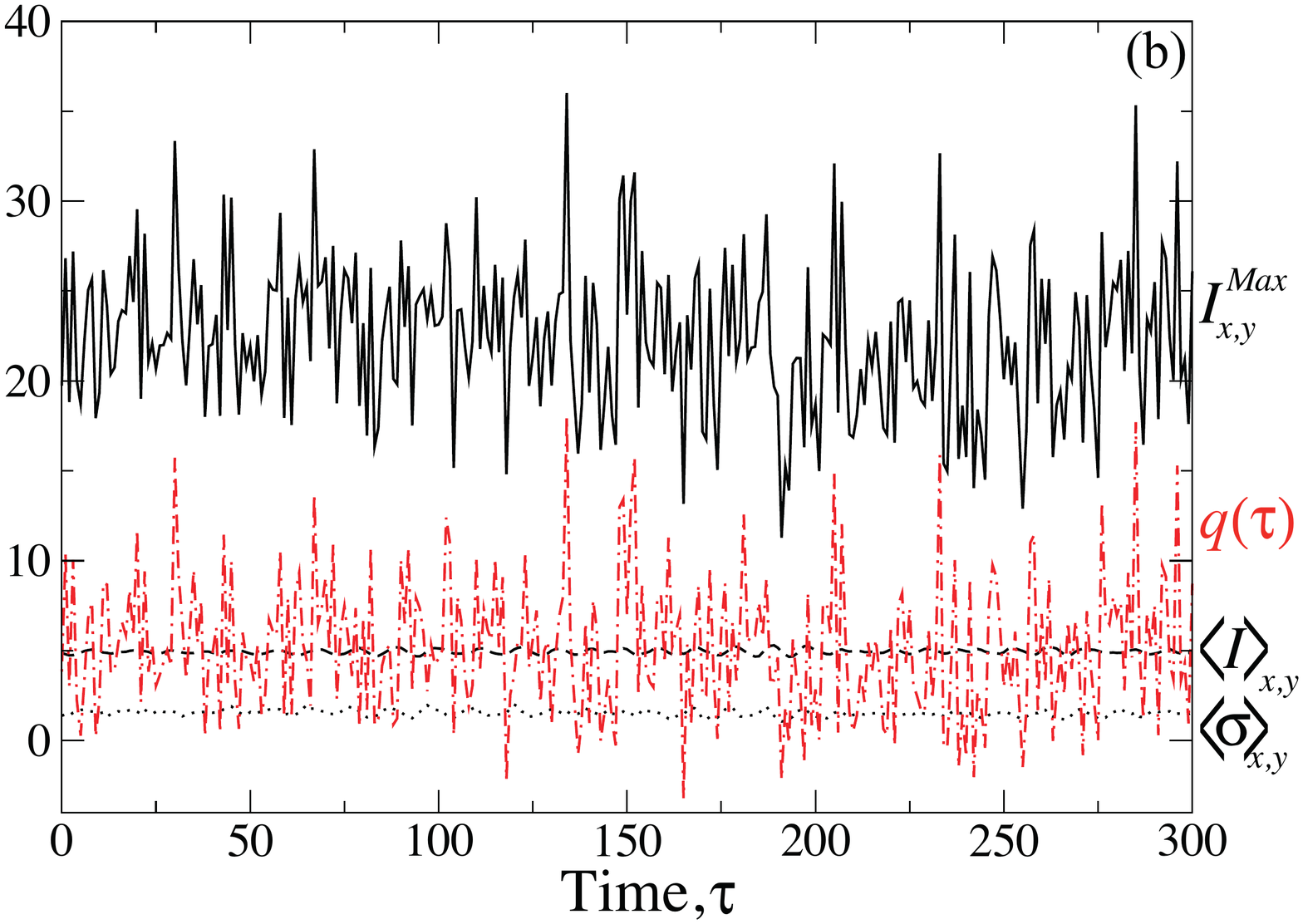} 
\caption{ \label{fig:turb2and8} (Color online) Temporal evolution of the maximum SROPO intensity (solid black 
line), the spatially averaged SROPO intensity (dashed black line), its standard deviation (dotted black line) 
and $q(\tau)$ for the sinc$^2$ model (\ref{SROPO}). Parameters are (a) $|E_0|^2=2$, $\theta=-0.3$, 
$|E_{IN}|^2=0.04$ and (b) $|E_0|^2=8$, $\theta=-1$, $|E_{IN}|^2=2.19$. All variables are dimensionless.}
\end{figure}

To characterize the regime of optical turbulence we have considered the temporal evolutions of the 
maximum SROPO intensity, the spatial average of the SROPO intensity and its standard deviation. As 
displayed in Figure \ref{fig:turb2and8}, the spatial statistics is large enough to guarantee probability 
distributions of well defined averages and deviations. Larger values of the pump power increase the size 
of the probability distribution of the SROPO intensity and that of the fluctuations of its maximum value 
(compare Figures \ref{fig:turb2and8}(a) and (b)). Such increase results in the occurrence and propagation 
of transverse rogue waves. 

Following the generally accepted definition of rogue waves in systems with injection \cite{jorge11_13}, we 
plot the temporal evolution of
\begin{equation}
\label{rwdef}
q(\tau) = I^{Max}_{x,y} (\tau) - \langle  \langle I \rangle_{x,y} \rangle_{\tau} - 
8 \langle  \langle  \sigma \rangle_{x,y} \rangle_{\tau} 
\end{equation}
corresponding to transverse pulse maxima, $I^{Max}_{x,y}$, above or below a threshold given by the 
average value of the intensity, $ \langle I \rangle_{x,y}$, plus eight times the standard deviation, 
$\sigma_{\tau,x,y}$, of the SROPO intensity for the sinc$^2$ model in the dashed-dotted red lines of 
Figures \ref{fig:turb2and8}(a) and (b). The presence of peaks of a rogue wave is signalled by positive 
values of $q(\tau)$ \cite{jorge11_13}. With pump intensities a few times above threshold (Figure 
\ref{fig:turb2and8}(a)), the rogue wave test fails ($q(\tau)$ 
remains negative) and the optical turbulence generated by the unlocking of the seed laser and the SROPO 
is relatively mild. With larger values of the pump power, however, rogue waves are commonplace and affect 
the spatio-temporal evolution of the SROPO field for long durations of the temporal evolution (see Figure 
\ref{fig:turb2and8}(b)). When comparing these results with those related to lasers with injections 
\cite{jorge11_13}, we note that our simulations are fully spatio-temporal and show that the material 
dynamics, typical of semiconductor media, is not essential in the generation and maintenance of rogue 
waves during optical turbulence. The main mechanism underlying rogue waves in SROPOs is the absence 
of locking between master and slave devices leading to intermittent phase jumps. Full investigations 
of optical turbulence in injected (seeded) optical devices will be presented elsewhere.

Finally, we have studied the presence and stability of cavity solitons (CS) in SROPOs with a particular 
focus on localised structures induced by the sinc$^2$ nonlinearity, i.e. away from threshold and with large 
seeding from an external laser. CS have been described in a variety of OPO devices without seeding from 
degenerate \cite{longhi97,staliunas97,oppo99,oppo01} to non-degenerate triply resonant configurations 
\cite{sanchez97,longhi98,devalcarcel00}. CS in degenerate OPOs have also been numerically extended 
to include the presence of seeding \cite{staliunasARXIV}. In the case of the non-degenerate SROPOs investigated 
here, the resonance condition of SROPO operation rules out any CS in the absence of seeding. It is then 
important to stress that all CS solutions described in this section are due to the external seeding field 
and have no counterpart in the case of $E_{IN}=0$.  

\begin{figure}[!tb]
\includegraphics[width=0.45\textwidth]{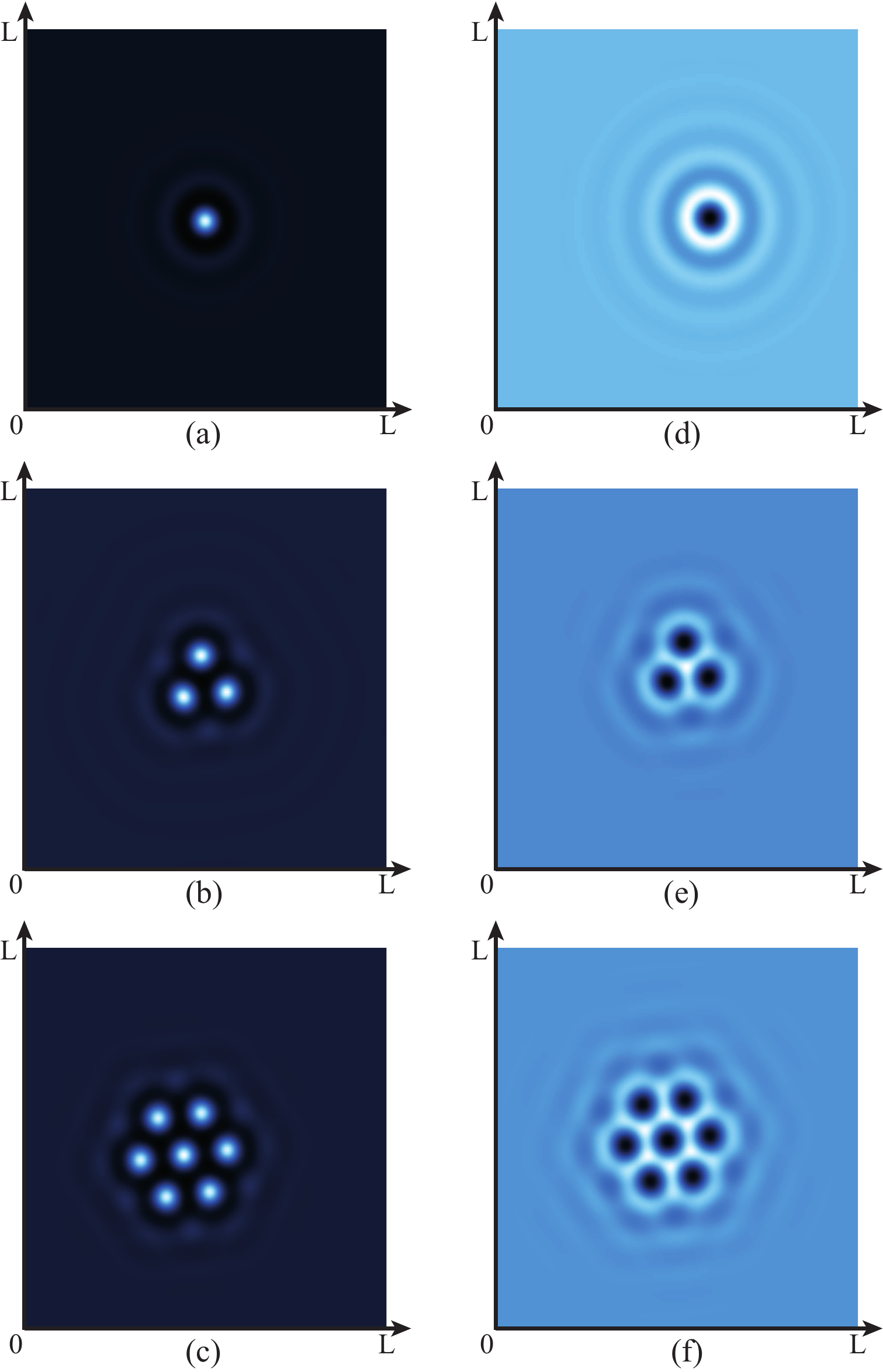} 
\caption{ \label{fig:CS} (Color online) Stable bright (a)-(c) and dark (d)-(f) CS configurations of the 
SROPO model (\ref{SROPO}). Parameters are $|E_0|^2=8$, $\theta=-1$, $|E_{IN}|^2=20.8$ for (a)-(c), 
$|E_{IN}|^2=26.5$ for (d) and $|E_{IN}|^2=27.1$ for (e)-(f). All variables are dimensionless.
}
\end{figure}

\begin{figure}[!b]
\includegraphics[width=0.5 \textwidth]{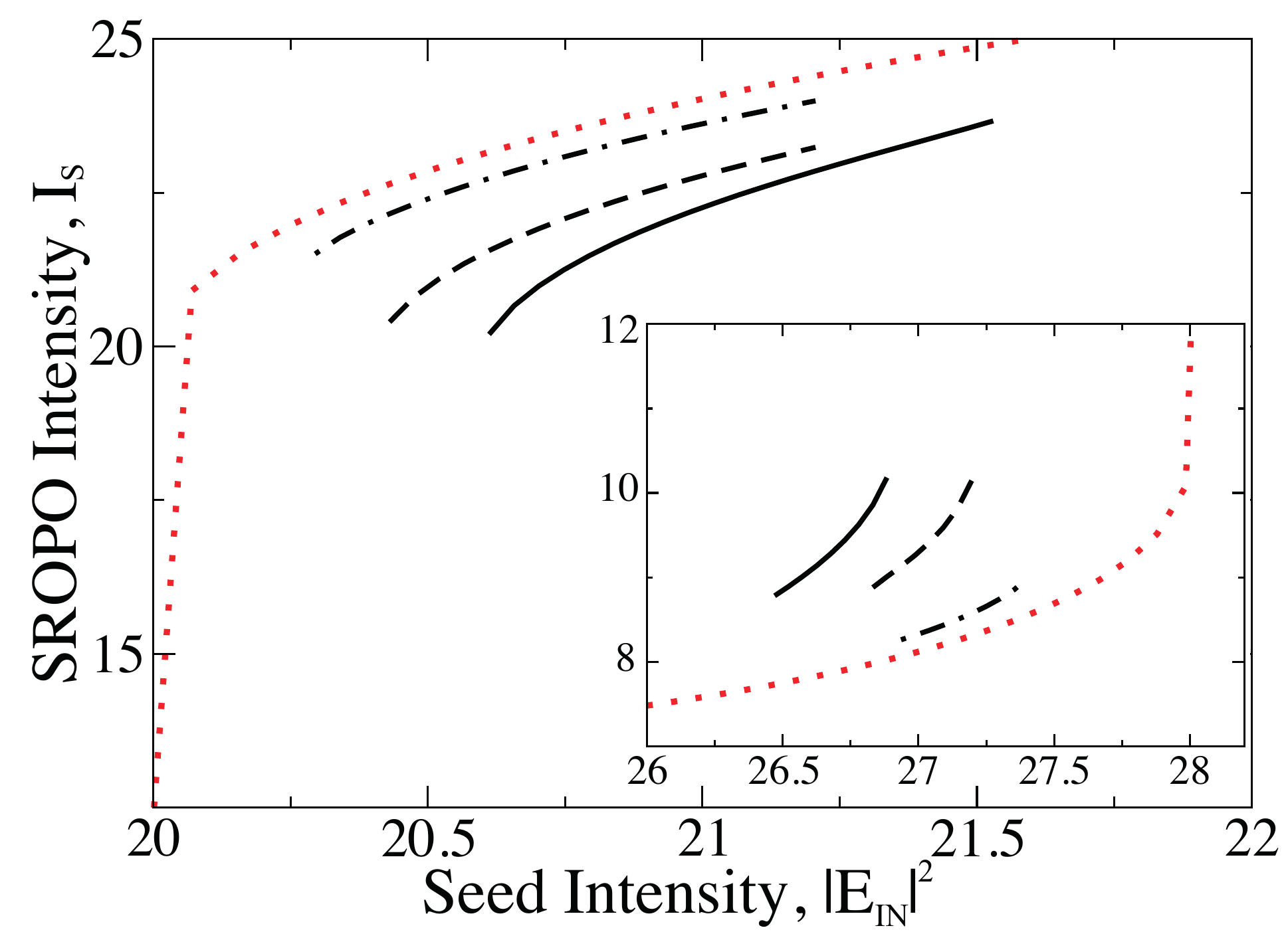} 
\caption{ \label{fig:csrange} 
(Color online) Stability range of clusters of CS for bright (main figure) and dark (inset) CS. 
Solid, dashed and dot-dashed black lines correspond to the maximum SROPO intensity in clusters 
of one, three and seven CS. The red dotted line corresponds to the maximum (minimum) intensity 
of the stable hexagonal (honeycomb) pattern in the main figure (inset). 
Parameters are $|E_0|^2=8$ and $\theta=-1$. All variables are dimensionless.
}
\end{figure}

Since we have introduced the sinc$^2$ nonlinearity in spatio-temporal models of SROPOs to describe 
self-organization when pump depletion and back--conversion take place, we focus here on CS in the limit 
of large pump powers. From Figures \ref{fig:PFe02_8} and \ref{fig:instabregime} we see that there are broad 
ranges of the parameter space where bistability between the plane wave solution and pattern structures is 
observed. For example, we find coexistent hexagons and homogeneous solutions for $|E_{IN}|^2$ between 19.98 
and 21.90 and coexistent honeycombs and homogeneous solutions for $|E_{IN}|^2$ between 26.00 and 28.09. Note 
that we even observe tri-stability among plane waves, hexagons and rolls for $|E_{IN}|^2$ between 21.25 and 
21.90 and among plane waves, honeycombs and rolls for $|E_{IN}|^2$ between 26.00 and 26.52. In the two 
wide regions of homogeneous-pattern bistability we have been able to locate single peak (bright) and single 
trough (dark) CS as shown for example in Figure \ref{fig:CS} (a) and (d), respectively. The onset and nature 
of these CS are again similar to those observed in nascent optical bistability \cite{tlidi94}. Together with 
the single unit bright and dark CS we have also found many multi-peak \cite{mcsloy02} and multi-trough localized 
structures that correspond to clusters of CS (also referred to as localized patterns \cite{lloyd08}). A few 
examples of these bright and dark clusters are displayed in Figure \ref{fig:CS}. The range of existence of 
single unit CS and CS clusters is displayed in Figure \ref{fig:csrange}. Snaking of both bright and dark CS 
is observed with stability branches of larger and larger clusters approaching the pattern stability lines 
in the parameter space (see Figure \ref{fig:csrange}). The details of the bifurcations and of the number of 
branches of bright and dark CS for changing $|E_0|^2$ are too long to be described here and will 
be the subject of a future publication. 

\section{\label{end} Conclusions.}
Self-organization and pattern formation in OPOs has been known for a number of years in degenerate 
\cite{oppo94a} or doubly or triply resonant non-degenerate configurations 
\cite{longhi96,sanchez97,santagiustina02}. The case of a widely non-degenerate SROPO has, however, 
been overlooked because of experimental limitations, now overcome, and the fact that off-resonance 
operation is inhibited because of its intrinsic tuneability. Here we have shown that under the action 
of a detuned injection close to the signal frequency, one can find an extremely rich variety of self-organized 
structures, from regular co-existing patterns to clusters of CS and even optical turbulence. In particular, 
we have derived mean field models for SROPOs with external seeding and shown that, away from threshold, 
cubic nonlinearities should be replaced by sinc$^2$ terms. The sinc$^2$ nonlinearity is capable of 
describing regimes of pump depletion and back-conversion. In these regimes, the external seeding generates 
hexagonal, roll and honeycomb patterns as well as bright and dark CS. Note that CS in SROPOs offer 
positional control associated to the generation of entangled photons with vastly different frequencies. 

In contrast to laser systems, the fast material dynamics of $\chi^{(2)}$ media makes a SROPO 
with external seeding an ideal candidate for comparisons between theory and experiments of 
optical self-organization. The fast material dynamics is also beneficial to the investigation of 
spatio-temporal structures in the regime of short pulse generation where many of the results presented 
here can find useful extensions. These investigations together with the full characterization of the 
turbulent regimes will be the subject of future communications.

\begin{center}
\textbf{Acknowledgments.} 
\par\end{center}

We thank J. Lega for useful discussions. We acknowledge financial support from the EU grant HIDEAS.
AMY acknowledges financial support from the UK Engineering and Physical Sciences Research Council (EPSRC).

\end{document}